\documentclass[twocolumn,aps,prl,superscriptaddress,showpacs,nofootinbib]{revtex4}
\usepackage{graphicx}
\usepackage{amsmath}
\usepackage{epsfig}
\usepackage{acronym}
\usepackage[utf8]{inputenc}
\usepackage[dvipdfm]{hyperref}

\newcommand{\ket}[1]{\ensuremath{\left\vert #1\right\rangle}}
\newcommand{\expval}[1]{\ensuremath{\left\langle #1 \right\rangle}}
\newcommand{\hsp}[1]{\hspace{#1 em}}
\newcommand{\sqz}{\hsp{-0.1}}
\newcommand{\ketbra}[2]{\left\vert{#1}\right\rangle \sqz\sqz\sqz \left\langle{#2}\right\vert}
\newcommand{\braket}[2]{\left\langle{#1}\right\vert \sqz \sqz \left. {#2}\right\rangle}

\begin{document}
 \title{Optimal working points for continuous-variable quantum channels}

\author{Imran Khan}
\affiliation{Max Planck Institute for the Science of Light, Guenther-Scharowsky-Str. 1/Bldg. 24, 91058 Erlangen Germany}
\affiliation{Institute for Optics, Information and Photonics, University of Erlangen-Nuernberg, Staudtstraße 7/B2, 91058 Erlangen, Germany}

\author{Christoffer Wittmann}
\affiliation{Max Planck Institute for the Science of Light, Guenther-Scharowsky-Str. 1/Bldg. 24, 91058 Erlangen Germany}
\affiliation{Institute for Optics, Information and Photonics, University of Erlangen-Nuernberg, Staudtstraße 7/B2, 91058 Erlangen, Germany}

\author{Nitin Jain}
\affiliation{Max Planck Institute for the Science of Light, Guenther-Scharowsky-Str. 1/Bldg. 24, 91058 Erlangen Germany}
\affiliation{Institute for Optics, Information and Photonics, University of Erlangen-Nuernberg, Staudtstraße 7/B2, 91058 Erlangen, Germany}

\author{Nathan Killoran}
\affiliation{Institute for Quantum Computing and Department of Physics \& Astronomy, University of Waterloo, N2L 3G1 Waterloo, Canada}

\author{Norbert Lütkenhaus}
\affiliation{Institute for Quantum Computing and Department of Physics \& Astronomy, University of Waterloo, N2L 3G1 Waterloo, Canada}

\author{Christoph Marquardt}
\affiliation{Max Planck Institute for the Science of Light, Guenther-Scharowsky-Str. 1/Bldg. 24, 91058 Erlangen Germany}
\affiliation{Institute for Optics, Information and Photonics, University of Erlangen-Nuernberg, Staudtstraße 7/B2, 91058 Erlangen, Germany}

\author{Gerd Leuchs}
\affiliation{Max Planck Institute for the Science of Light, Guenther-Scharowsky-Str. 1/Bldg. 24, 91058 Erlangen Germany}
\affiliation{Institute for Optics, Information and Photonics, University of Erlangen-Nuernberg, Staudtstraße 7/B2, 91058 Erlangen, Germany}

\date{\today}

\begin{abstract}

The most important ability of a quantum channel is to preserve the quantum properties of transmitted quantum states. We experimentally demonstrate a continuous-variable system for efficient benchmarking of quantum channels. We probe the tested quantum channels for a wide range of experimental parameters such as amplitude, phase noise and channel lengths up to 40\,km. The data is analyzed using the framework of effective entanglement. We subsequently are able to deduce an optimal point of operation for each quantum channel with respect to the rate of distributed entanglement. This procedure is a promising candidate for benchmarking quantum nodes and individual links in large quantum networks of different physical implementations.

\end{abstract}


\maketitle
It is envisaged that a quantum network would consist of many nodes and channels for processing and distribution of information \cite{Kimble2008}. Such a network requires the ability to perform well-controlled operations on single quantum systems and to generate entanglement between several quantum systems. In the past, considerable efforts were made to implement, characterize and optimize such local operations \cite{Hanneke2009}. As the complexity of single nodes increases, another crucial aspect to consider is the transfer of quantum information within and between them. In quantum networks, as well as quantum computers, characterization of the individual nodes and channels via process tomography \cite{Sanders_Lvovsky_2008} or a direct benchmarking procedure is essential to get a performance estimate of the whole system \cite{Haseler2010, Killoran2011}. Moreover, if we can analyze nodes and channels \emph{quantitatively}, then, by tuning parameters, we can optimize their operational working points.

A common way to evaluate the performance of quantum devices is to compare them against analogous classical devices using fidelity-based benchmarks \cite{Braunstein_Fuchs_Kimble_2000}, even though this approach can be experimentally costly. While fidelity may allow us to distinguish devices from their classical counterparts, it is not clear how to use it to quantitatively characterize devices. Results from the field of \ac{QKD} can provide another possible way to quantitatively characterize quantum networks. In \ac{QKD}, one aims to generate a secret key from quantum-correlated data by measuring quantum states shared between distant parties. The quantum states are exchanged over a quantum channel. This channel is an important component and requires great care when being established and characterized \cite{Jain2010}. However, there are other less demanding applications for quantum channels, and the specific requirements for the channel will depend on our needs. This allows us to identify some minimal requirements \cite{Curty_Lewenstein_Luetkenhaus_2004}, based on the preservation of entanglement, for a channel to be useful for quantum communication.

From a more general perspective, the distribution of quantum correlations in quantum networks can be efficiently investigated using the framework of effective entanglement~\cite{rigas06a, Haseler2010, Killoran2011}. Although our approach applies generally, we focus here on continuous-variable systems, such as systems based on quantum optics, collective spin excitations \cite{Hammerer2010}, surface plasmons \cite{Huck2009}, optomechanics \cite{Schmidt2012} and cavity quantum electrodynamics \cite{Fedorov2012}. In contrast to fidelity-based benchmarks, the framework of effective entanglement aims to use minimal experimental resources while providing a quantitative characterization of quantum devices. So far, experiments have been performed to witness effective entanglement in a fiber \cite{Wittmann2010a} and a free-space \cite{Lorenz2006} channel, as well as a benchmark performed on a quantum memory \cite{Killoran2012}. It is now desireable to have a benchmark system that is capable of covering a wide range of parameters in order to find the optimal point of operation for each quantum device.

In this Letter, we demonstrate an experimental implementation of a continuous-variable system for efficient benchmarking of quantum channels and devices. This benchmark system allows us to identify the optimal point of operation of a given quantum channel with respect to the rate of transmitted entanglement. We find the optimum by varying the quantum state amplitude and measuring the excess noise for fiber channels with a length of up to 40\,km. The entanglement rates are calculated using the framework of effective entanglement. Furthermore, we demonstrate the impact of phase noise on the ability to distribute entanglement, showing that after a certain amount of dephasing the detectable quantum correlations drop to zero. 

The primary goal of quantum benchmarking is to operationally differentiate a given channel/device from the class of measure and prepare (MP) channels. MP channels involve first a measurement, followed by classical storage/transmission of the measurement result, and finally the preparation of output states based on this classical data. Due to the classical step, MP channels are not sufficient for quantum communication protocols. Accordingly, any channel which cannot be distinguished from MP operation will not be usable in a quantum communication context. 

The set of MP channels is mathematically equivalent to the set of entanglement-breaking channels \cite{horodecki03a}. Thus, if we show that a channel can preserve entanglement, then that channel has passed our quantum benchmark. In fact, by using a source-replacement (or prepare and measure) description, the entanglement-preserving capabilities of a channel can be efficiently tested without needing to use real entangled states. In such a scheme, we probe the channel with different test states (for the experiment, we use two coherent states $\ket{\pm\alpha}$). Theoretically, we imagine that these test states are part of a larger entangled state, given by \cite{rigas06a,haseler08a}
\begin{equation}
 \ket{\Psi}_{AA'}=\frac{1}{\sqrt{2}}\left[\ket{0}_A\ket{\alpha}_{A'}+\ket{1}_A\ket{-\alpha}_{A'}\right].
\end{equation}
We represent the tested channel mathematically by a quantum map $\Lambda$, mapping from the input system $A'$ to some output system $B$. To test the entanglement-preserving capabilities of the channel, we need to determine whether or not the bipartite output state $\rho^\mathrm{out}_{AB}=(\mathrm{id}\otimes\Lambda)\left[\ketbra{\Psi}{\Psi}_{AA'}\right]$ is entangled. However, since the entangled state is a virtual construction, we do not have access to tomographically-complete information of this effectively entangled state. Our information is limited to only the reduced density matrix $\rho_A=\mathrm{Tr}_B(\rho_{AB}^\mathrm{out})$, along with any measurements made on the physical output states $\rho_{0/1}^\mathrm{out}$. The reduced density matrix
\begin{equation}
 \rho_A = \frac{1}{2}\begin{bmatrix}
                      1 & s\\
		      s^* & 1
                     \end{bmatrix},
\end{equation}
is parameterized by the overlap $s = \braket{\alpha}{-\alpha}$ and remains constant during the protocol. Previously, we have used Stokes measurements \cite{Wittmann2010a} because they enable benchmarking under the adversarial scenario common in \ac{QKD}. Here, we are interested in quantifying channels under less restrictive conditions. In this case, the conjugate quadrature operators $\hat{x}$ and $\hat{p}$ suffice (see \cite{haseler08a} for discussion about Stokes vs quadrature measurements). Through measurement, we determine the mean value $\expval{\hat{z}}$ and variance $\mathrm{Var}(\hat{z})=\expval{\hat{z}^2}-\expval{\hat{z}}^2$ of $\hat{z}\in\{\hat{x},\hat{p}\}$ for each output state. 

Using the overlap value and the obtained measurement results, the task is to determine whether $\rho^\mathrm{out}_{AB}$ is entangled. One method for accomplishing this is with a witness-like procedure \cite{rigas06a,haseler08a}. Alternatively, we can minimize an appropriate entanglement measure over all bipartite states consistent with the available information \cite{killoran10b, Killoran2011}. In addition to classifying entangled states \cite{Killoran2011}, the entanglement measure approach gives us the ability to also quantitatively describe and compare different channels. Although we use the concept of effective entanglement to simplify our benchmarking scheme, we note that it is entirely representative of what one would find had the same data been generated from a real entangled state.

Our entanglement measure cannot be chosen arbitrarily; the quantification procedure outlined in \cite{killoran10b,Killoran2011} requires that the measure satisfies some mild monotonicity properties. In addition, the measure should ideally be efficiently computable, for instance via a semidefinite program. In this paper, we will use the negativity \cite{zyczkowski98a,lee00a,vidal02a}, which fulfills the required conditions. It will also be convenient to convert our entanglement content to the logarithmic negativity \cite{plenio05a}, another measure which is in one-to-one correspondence with the negativity. The log-negativity has the useful property of being additive, allowing us to quantify entanglement built up over repeated channel use. It also has an operational interpretation in terms of the entanglement cost under a certain class of operations \cite{Audenaert2003}.
\begin{figure}
	\begin{tabular}{l}
\includegraphics[width=0.4\textwidth]{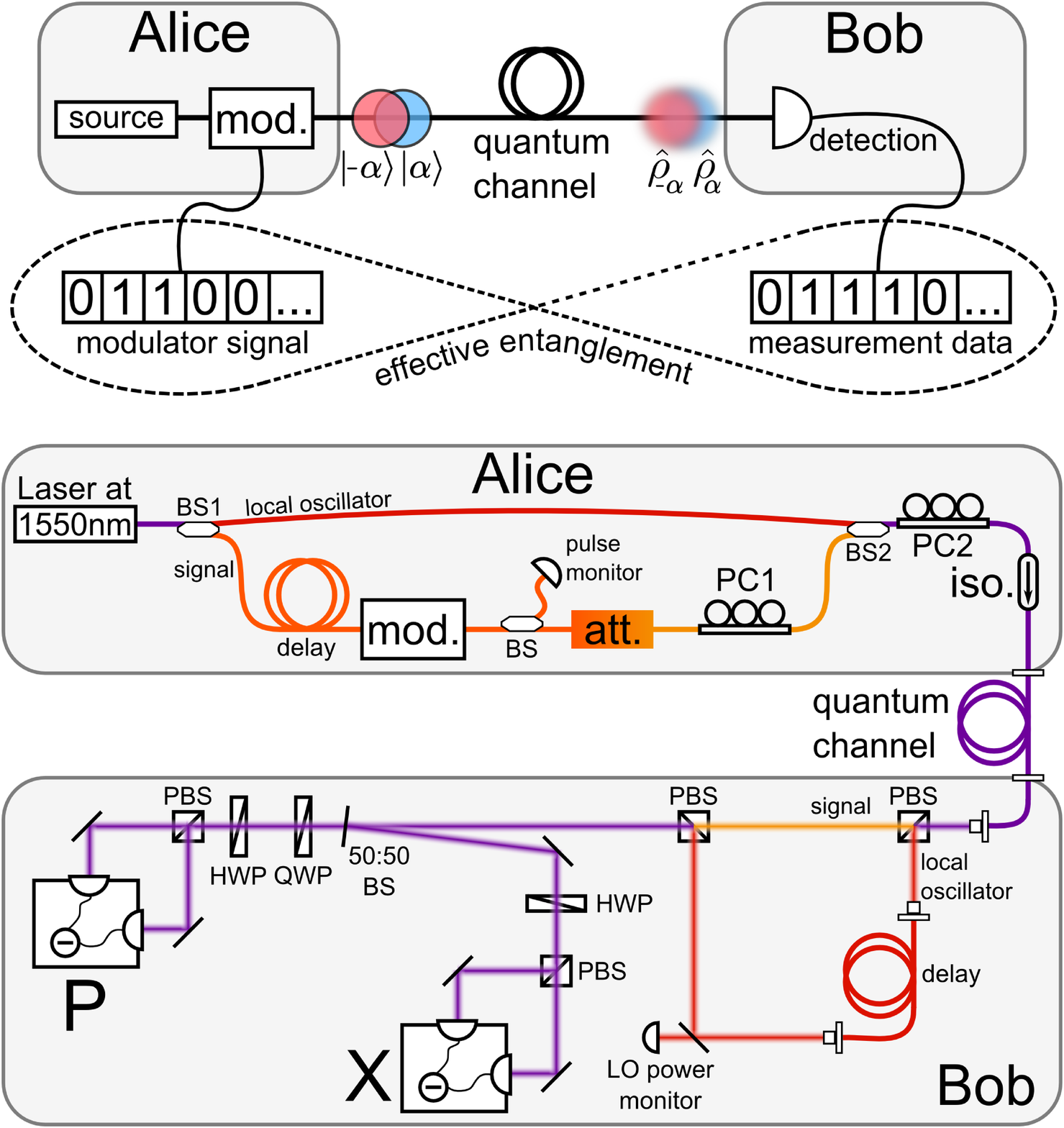}\\[-7.7cm]
\hspace{-.5cm}(a) \\[2.5cm]
\hspace{-.5cm}(b)\\[4.5cm]
\end{tabular}
	\caption{(a) Quantum benchmarking system using a prepare-and-measure-type setup. Effective entanglement can be quantified using Bob's measurement data of the signal states. (b) Schematic drawing of the experimental setup. The abbreviated components are  beamsplitters (BS), the Mach-Zehnder modulator (mod.), an attenuator (att.), an optical isolator (iso.), a polarizing beamsplitter (PBS), quarter-wave plate (QWP) and half-wave plate (HWP).}\label{fig:setup}
\end{figure}

We implemented the benchmark using an experimental setup similar to that demonstrated in \cite{Wittmann2010a}, consisting of three parts: a sender (Alice), a receiver (Bob) and the channel to be tested (which is simply ``plugged in'' to the benchmark system) (Fig. \ref{fig:setup}). Alice and the quantum channel are fiber-based, whereas Bob uses free space optics. We use a 1550 nm laser as a light source, operated at a repetition rate of 1 MHz, producing pulses of 20ns. The pulses are then split up asymmetrically (BS1) into a \ac{LO} and a signal line. The larger portion of the beam is directed to the \ac{LO} line to provide a strong phase reference for the weak signal beam. The signal preparation line is composed of a delay fiber, a \ac{MZM}, an attenuator and a polarization controller. The delay fiber causes a 0.5 $\mu s$ delay between signal and \ac{LO} pulses to prevent scattering of photons from the \ac{LO} into the signal mode during their propagation through the channel. The \ac{MZM} generates coherent states with opposite phases $0$ and $\pi$ while the attenuator reduces the state amplitude to a quantum level. A polarization controller (PC1) prepares the signal states' polarization orthogonal to the \ac{LO} pulses. After recombining the two modes with a beamsplitter (BS2) another polarization controller (PC2) is adjusted such that \ac{LO} and signal can be split up again on Bob's side. The two modes are then sent through an isolator to the quantum channel, after which they enter Bob's setup. There, the time multiplexing delay is reversed to allow for temporal mode matching of signal and \ac{LO}. The signal states are then characterized using double homodyne detection. Additionally, the signal pulses in Alice and the \ac{LO} in Bob are monitored.

We employ a binary alphabet to probe the quantum channel. The alphabet consists of two weak coherent states: $|\alpha\rangle$ and $|\mathrm{-}\alpha\rangle$ ($|\alpha| \approx 0.5)$. This alphabet is chosen, because it is easy to implement (single \ac{MZM}) and to post-process. After the states have passed through the quantum channel they enter Bob's receiver. We measured a receiver efficiency $\eta_{Bob}$ of $\approx 73\%$, which consists of the optical losses in Bob's setup, the homodyne detector efficiencies and the interferometric visibility between the signal and \ac{LO} mode. Note that the total transmission $T$ used later on includes the receiver efficiency $\eta_{Bob}$ and the losses in the quantum channel. Bob's double homodyne detection measures the joint probability distribution of the conjugate quadrature operators $\hat{x}$ and $\hat{p}$, also known as the Q-function \cite{Stenholm_1992}. We obtain the marginal distribution of the Q-function for each quadrature to calculate its first and second moments. From the first moments we compute the complex number $\beta = \langle\hat{a}\rangle$ \footnote{$\hat{a}$ denotes the quantum mechanical annihilation operator for the light mode under investigation.} for which we use the symmetric definition $\langle\hat{a}\rangle = \frac{1}{\sqrt{2}}(\langle\hat{x}\rangle + i \langle\hat{p}\rangle)$. Knowing the total transmission $T$ we are able to calculate the overlap $\braket{\alpha}{-\alpha}$ of the coherent states on Alice's side.  The second moments of the Q-function marginals, minus a constant, correspond to the quadrature variances Var($\hat{x}$) and Var($\hat{p}$). We normalize the measured variances of our signal states to that of the coherent vacuum state, where we chose the shot noise level to be equal to 1. Any additional variance is considered to be excess noise added by the channel.

\begin{figure}
	\includegraphics[width=0.5\textwidth]{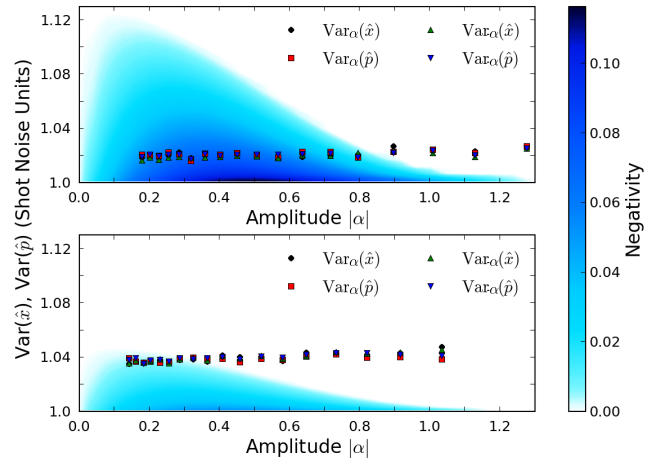}
	\caption{Quantification of effective entanglement for a 20 km (top) and a 40 km (bottom) fiber channel. The negativity values in the shaded region were calculated by assuming all variances equal \cite{Killoran2011}. Since the experimental variances have small differences in $\hat{x}$ and $\hat{p}$ of the alphabet, the theoretical negativity values here serve only as a reference. We observe that the 20 km channel can tolerate more excess noise whereas the 40 km channel shows to be at its limit. Note that the displayed negativity is a lower bound to the actual transmitted entanglement and that the statistical error is contained within the data points.}
	\label{fig:eeattseries}
\end{figure}
\begin{figure}
	\includegraphics[width=0.5\textwidth]{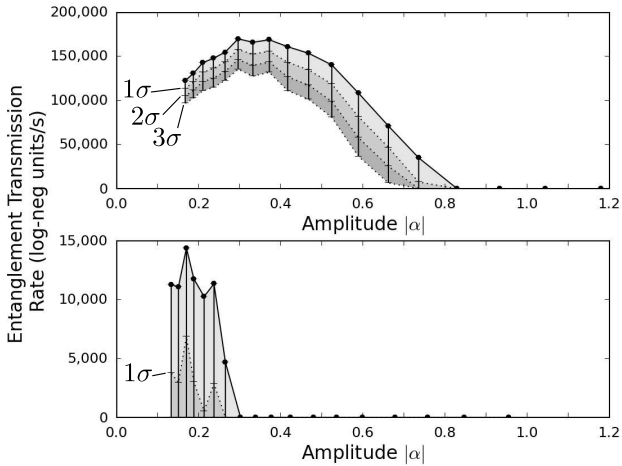}
	\caption{Minimal rate of distributed entanglement for a 20 km (top) and a 40 km (bottom) channel, obtained using experimentally measured variances. Temporal variability in the entanglement transmission rates are determined by including 1$\sigma$, 2$\sigma$ and 3$\sigma$ error bars in the optimization. In the experiment the temporal variability corresponds to fluctuations in the channel.}
	\label{fig:eeattseries2}
\end{figure}

The parameters we varied are the transmission of our quantum channel, the sending amplitude $|\alpha|$ of the coherent test states and the magnitude of the artificially induced phase noise. We performed the benchmark for two standard fibers with lengths of 20 km ($T\approx\!24\%$) and 40 km ($T\approx\!9\%$) serving as the quantum channel. We probed the channels for different sending amplitudes ranging approximately from  $0 \leq |\alpha| \leq 1$ and found a nearly constant quadrature variance of $\approx\!1.02$ (20 km fiber) and $\approx\!1.04$ (40 km fiber). We were able to witness and quantify the preservation of entanglement for both quantum channels for certain amplitudes (Fig. \ref{fig:eeattseries}). Using a sending rate of 0.875~MHz and the quantified log-negativity we calculated the rate of distributed entanglement for both channels (Fig. \ref{fig:eeattseries2}). We find maximum rates of 166,000 log-neg units/s for a 20 km and 15,000 log-neg units/s for a 40 km channel. For comparison, a maximally-entangled two-qubit state has 1 log-neg unit of entanglement. We were thus able to determine the optimal point of operation for the two quantum channels (Table \ref{tab:ee-results}).

\begin{table}
\caption{\label{tab:ee-results}Deduced optimal point of operation for the tested fiber quantum channels.}
\begin{ruledtabular}
\begin{tabular}{ccc||ccccc}
 &T&\;$|\alpha|$\;\;&$\sim\text{Var}(\hat{x}, \hat{p})$&negativity&log-neg&rate\\
\hline
20 km& 24\% & 0.3 \;& 1.02 &0.07&0.19&166,000\\
40 km& 9\% & 0.17 \;& 1.04 &0.006&0.017&15,000\\
\end{tabular}
\end{ruledtabular}
\end{table}

Furthermore, we investigated the impact of excess noise on the channel and the effective entanglement benchmarking protocol. Here, our phase calibration procedure provides access to a \textit{tunable} noise parameter. As no phase lock is present between signal and \ac{LO} pulses, their relative phase drifts freely whenever they propagate through different spatial paths of the setup. Without any post-processing the signal states thus arrive effectively phase-randomized. However, we experimentally find that their relative phase remains fairly constant during a 250 $\mu s$ time frame. Alice therefore sends along bright calibration pulses to provide a classical phase reference for Bob. These calibration pulses are used to rotate the signal pulses in the measured phase space to match Alice's preparation frame of reference. This procedure effectively reverts the phase-randomization \cite{Qi_Huang_Qian_Lo_2007}.

\begin{figure}
	\includegraphics[width=0.5\textwidth]{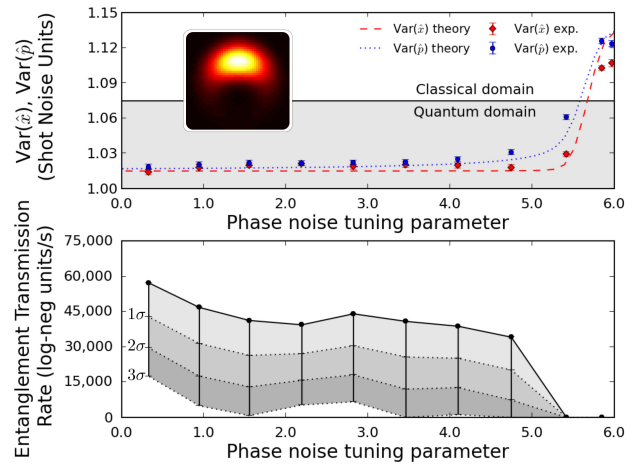}
	\caption{Examination of entanglement throughput versus phase noise (the phase noise tuning parameter is defined as $6 - |\alpha_{\text{cal}}|$). The transition between quantum and classical domains can be observed for increasing phase noise (diminishing calibration amplitude). This is reflected in increased quadrature variances (top) and subsequently decreasing transmission rates (bottom). The rates have been obtained by computing the log-negativity for the measured asymmetric variances. The channel was probed for $|\alpha| \approx 0.5$. (Inset figure: Illustration of a phase-diffused coherent state.)}
	\label{fig:cpa}
\end{figure}

The phase noise may now be tuned, by gradually weakening the calibration pulses and thus diminishing the phase information available for the phase space rotation. We therefore expect a gradual phase-diffusion of the signal states with decreasing calibration pulse amplitude. This phase-diffusion is reflected by an increased quadrature variance of $\hat{x}$ and $\hat{p}$. Below a certain calibration pulse amplitude, the signal states will be phase-diffused to such a degree, that the negativity drops to zero and entanglement can no longer be witnessed. Fig. \ref{fig:cpa} shows this effect for a 20 km fiber channel.

The theoretical curves for Fig. \ref{fig:cpa} were obtained by numerically calculating the variance of the Q-function marginals of phase-diffused coherent states. The phase-diffused Q-function $Q_{pd}$ was obtained by calculating the weighted integral of Q-functions of coherent states with the same amplitude $r$ but different phase:
\begin{equation}
Q_\mathrm{pd}(\beta) = \frac{1}{\pi}\int_{\phi=0}^{2\pi}f(\phi)\left|\braket{r e^{i\phi}}{\beta}\right|^2 \mathrm{d}\phi.
\end{equation}

The amount of phase-diffusion of $Q_{pd}$ is determined by the phase probability distribution $f(\phi)$ of a coherent state, given by $f(\phi) = \int Q_{\mathrm{coh}}(r,\phi)~r\mathrm{d}r$. The amplitude of $Q_{\mathrm{coh}}$ is equal to the amplitude of the calibration states Bob receives, thus corresponding to the amount of phase information available to him. The asymmetry between Var$(\hat{x})$ and Var$(\hat{p})$ is merely a matter of choosing a fixed frame of reference in phase space for the simulated phase-diffused state (modulation along $\hat{x}$).

In conclusion, we demonstrated a versatile continuous-variable system for efficient benchmarking of quantum communication channels and devices. This procedure enables us to investigate the quantum throughput of various quantum systems independent of their physical realization. We can probe the channel over a range of tunable parameters, such as input state amplitude and phase noise, allowing us to determine optimal working points. Using the framework of effective entanglement, we determined these points with respect to the rate of distributed entanglement.

This work was supported by the BMBF (Bundesministerium für Bildung und Forschung) grant for QuOReP (Quantenoptische Repeater-Plattform) and EU grant Q-ESSENCE. We thank Tobias Röthlingshöfer, Christoph Bersch and Georgy Onishchukov for technical support. NK is grateful for support from the Collaborative Student Training in Quantum Information Processing program and the Ontario Graduate Scholarship program.

\bibliography{literature}

\acrodef{LO}[LO]{local oscillator}
\acrodef{QKD}[QKD]{quantum key distribution}
\acrodef{BS}[BS]{beam splitter}
\acrodef{PBS}[PBS]{polarization beam splitter}
\acrodef{HWP}[HWP]{half-wave plate}
\acrodef{QWP}[QWP]{quarter-wave plate}
\acrodef{ebits}[ebits]{entanglement bits}
\acrodef{MZM}[MZM]{Mach-Zehnder modulator}
	
\end{document}